%% file: main.tex
\documentclass[11pt]{article}

\usepackage[margin=1in]{geometry}
\usepackage[T1]{fontenc}
\usepackage[utf8]{inputenc}
\usepackage{lmodern}
\usepackage{microtype}
\usepackage{booktabs}
\usepackage{array}
\usepackage{tabularx}
\usepackage{longtable}
\usepackage[backend=biber,style=numeric,sorting=none]{biblatex}
\usepackage{hyperref}
\usepackage{xurl}
\usepackage{setspace}
\usepackage{graphicx}
\usepackage{tikz}
\usetikzlibrary{arrows.meta,positioning,shapes.geometric,fit,backgrounds}

\hypersetup{
    colorlinks=true,
    linkcolor=blue,
    citecolor=blue,
    urlcolor=blue
}

\title{Governing Generative AI Across Financial Institutions: A Framework for Generative AI Risk Control}
\author{}
\date{}

\begin{document}

\begin{titlepage}
\centering
\maketitle

\vspace{0.8cm}

{\large
Dennis Mao \textsuperscript{2},\quad
Alessandra Lin\textsuperscript{1},\quad
Yixin Kang\textsuperscript{2},\quad
Yiqing Wang \textsuperscript{3}

\textsuperscript{1}\textit{Independent Researcher}, Dallas, TX, USA \\[3pt]
\textsuperscript{2}\textit{Florida State University}, Tallahassee, FL, USA \\[3pt]
\textsuperscript{3}\textit{Independent Researcher}, Chicago, IL, USA
}

\vspace{0.9cm}
\begin{abstract}
Generative artificial intelligence is moving from general-purpose experimentation toward specialized applications across banking, capital markets, insurance, payments, and wealth management. Its main contribution is not limited to conversational interfaces. Modern generative systems can synthesize large document collections, extract information from unstructured data, generate software and analytical code, create scenario narratives, support research workflows, and coordinate multi-step tasks. These capabilities make generative AI especially relevant to finance, where decisions often depend on combining quantitative data with contracts, policies, filings, news, customer communications, and expert judgment.

This paper presents an application-oriented view of generative AI in finance. It organizes potential uses around five capability patterns, including knowledge synthesis, content generation, analytical assistance, interaction, and workflow orchestration, and maps them to major financial functions. Representative applications include investment research, customer service, lending support, fraud investigation, financial reporting, operations automation, software development, and personalized financial guidance. The paper also discusses common technical architectures, such as retrieval-augmented generation, tool-using assistants, multimodal models, and agentic workflows, and identifies practical factors that shape business value. The resulting landscape provides a foundation for researchers and practitioners seeking to understand where generative AI may produce the greatest operational and analytical impact in financial services.
\end{abstract}


\end{titlepage}

\input{NewSections/01_introduction}
\input{NewSections/02_Context}
\input{NewSections/03_CLCF}
\input{NewSections/04_Conclusion}

\printbibliography

\end{document}

%% file: NewSections/01_introduction.tex
\section{Introduction}
\label{sec:introduction}

Financial institutions generate and consume large volumes of both structured and unstructured information. Prices, exposures, transaction histories, and risk measures coexist with annual reports, contracts, policies, research notes, customer messages, call transcripts, and regulatory publications. Traditional analytical systems are effective when the task is well specified and the data are structured, but they are less flexible when users must search, interpret, summarize, or combine information expressed in natural language. Generative AI expands the set of activities that can be supported computationally by allowing systems to create, transform, and reason over text, code, images, tables, and other forms of content \cite{bommasani2021foundationmodels,nie2024survey}.

Large language models can summarize long financial documents, compare disclosures across firms, draft research notes, translate technical material for different audiences, generate code for data analysis, and interact with external tools. When connected to enterprise data through retrieval-augmented generation (RAG), an object can answer questions using a selected collection of internal or external documents rather than relying only on its pretrained parameters \cite{lewis2020rag}. Multimodal models can jointly process text, charts, scanned forms, and images, while tool-using and agentic systems can execute a sequence of search, calculation, drafting, and review tasks \cite{okpala2025agentic,wu2023unleashing}.

These capabilities are particularly useful in finance because many workflows combine numerical analysis with interpretation and communication. An equity analyst may review earnings releases, transcripts, market data, and prior forecasts before updating an investment view. A lending specialist may combine application information, financial statements, and supporting documents. An operations team may reconcile transactions while interpreting exceptions described in free text. A customer-service representative may need to locate product terms and explain them clearly. In each case, generative AI can serve as a bridge between data, documents, software, and human users.

This paper develops an application-oriented landscape of generative AI in financial services. Rather than proposing a control or regulatory framework, it focuses on what generative systems may do, where they may be deployed, and how their technical capabilities translate into business use. Section~\ref{sec:capabilities} describes the core capability patterns and enabling architectures. Section~\ref{sec:applications} maps these capabilities to representative financial functions and use cases. Section~\ref{sec:conclusion} summarizes the main opportunities and future research directions.

%% file: NewSections/02_Context.tex
\section{Core Capabilities }
\label{sec:capabilities}

Generative AI applications in finance can be illustrated through some usage case examples.

\subsection{Knowledge Synthesis}

Knowledge synthesis can refer to searching, comparing, and summarizing information distributed across multiple sources. In finance, relevant evidence is often fragmented across filings, research reports, contracts, policies, emails, transcripts, and market commentary. A generative AI can produce concise summaries, extract key words, identify changes between document versions, and answer natural-language questions over a document collection as a chat bot. RAG is especially useful for this pattern because retrieved passages can provide current, domain-specific context to the model \cite{lewis2020rag,chen2026doesragknowretrieval}.

Typical outputs include earnings-call summaries, covenant extraction, product comparisons, policy digests, due-diligence briefs, and cross-document thematic analyses. The main productivity gain comes from reducing the time required to locate and organize information before an expert performs deeper analysis.

\subsection{Content and Communication Generation}

Generative models can transform technical or fragmented inputs into coherent communication. Examples include drafting client emails, preparing meeting summaries, producing first versions of research notes, translating financial explanations into plain language, and creating narrative commentary for dashboards or management reports. The output can be customized according to the audience, tone, length, and format \cite{dillon2025shifting}.

This capability can also support document standardization. An object can convert analyst notes into a common template, transform bullet points into a structured memorandum, or generate alternative versions of the same explanation for internal specialists and external customers. The value is greatest where users repeatedly convert similar evidence into different communication formats.

\subsection{Analytical and Coding Assistance}

Generative AI can help users formulate analytical questions, write code, interpret statistical output, and explain quantitative results. In financial data science, a model may generate SQL queries, Python scripts, spreadsheet formulas, test cases, or documentation. It may also explain model outputs, compare forecasting approaches, or convert a natural-language request into a sequence of calculations \cite{chen2025generative}.

This pattern does not replace numerical engines. Instead, the generative model acts as an interface to databases, statistical libraries, pricing systems, and visualization tools. The strongest designs separate language generation from deterministic calculation: the model interprets the task and selects tools, while specialized software performs the computation.

\subsection{Interactive Assistance}

Conversational systems allow users to explore financial information iteratively. A customer may ask follow-up questions about a transaction, an advisor may refine a portfolio screen, or an employee may query an internal knowledge base \cite{sai2025generative}.

Interactive assistance can be embedded in websites, mobile applications, contact centers, employee portals, and analytical workbenches. It can also support multilingual interaction, which is useful for global institutions and diverse customer populations.

\subsection{Workflow Orchestration}

Tool-using and agentic systems extend generation into multi-step execution. A workflow may retrieve documents, extract fields, run calculations, compare results, draft a response, and update a work queue. Agentic architectures are therefore relevant to processes that cross several systems or require repeated handoffs \cite{okpala2025agentic,wang2026safeskillscollidemeasuring}.

The practical distinction between an assistant and an agent is the degree of task completion. An assistant primarily recommends or drafts; an agent performs a sequence of actions toward an objective. In finance, early applications are likely to focus on bounded workflows such as document preparation, research compilation, exception investigation, and software maintenance.

\subsection{Common Technical Patterns}

Table~\ref{tab:technical_patterns} summarizes several architectures frequently used to build finance-oriented generative AI applications.

\begin{table}[htbp]
\centering
\small
\caption{Common technical patterns for generative AI in finance.}
\label{tab:technical_patterns}
\begin{tabularx}{\textwidth}{p{0.23\textwidth} X X}
\toprule
\textbf{Pattern} & \textbf{Description} & \textbf{Representative financial uses} \\
\midrule
Prompt-based assistant & A general or domain-adapted model responds to user instructions without direct access to private systems. & Drafting, rewriting, brainstorming, coding help, and general financial education. \\
\addlinespace
Retrieval-augmented generation & The model receives passages retrieved from a selected knowledge base. & Research search, contract review, policy Q\&A, product information, and document comparison. \\
\addlinespace
Tool-using copilot & The model calls databases, calculators, pricing engines, or analytical libraries. & Portfolio analysis, financial modeling, transaction investigation, and reporting. \\
\addlinespace
Multimodal document intelligence & The system processes text together with tables, charts, forms, and scanned images. & Financial statement spreading, invoice processing, claims analysis, and KYC document review. \\
\addlinespace
Agentic workflow & Multiple steps are planned and executed across tools or specialized agents. & Due diligence, research pipelines, exception handling, software testing, and operations automation. \\
\bottomrule
\end{tabularx}
\end{table}

%% file: NewSections/03_CLCF.tex
\section{Generative AI Applications Across Finance}
\label{sec:applications}

The potential uses of generative AI span the financial value chain. Figure~\ref{fig:application_landscape} presents a capability-to-application map that differs from a sequential process model: core generative capabilities form the center, while major financial domains surround them.

\subsection{Investment Research and Market Intelligence}

Investment research is a natural application because analysts routinely combine structured market data with unstructured information. Generative AI can summarize earnings releases and conference calls, compare current disclosures with prior periods, extract management guidance, identify recurring themes, and draft initial research notes. It can also create question lists before management meetings or organize evidence supporting competing investment theses \cite{nie2024survey,bagattini2025leveraging}.

When linked to market-data and analytical tools, a research copilot may move from document summarization to interactive analysis. For example, a user could request a comparison of revenue growth, valuation multiples, and management sentiment across a peer group. The system could retrieve filings, calculate metrics, and produce a narrative explanation. Similar methods can support macroeconomic research by synthesizing central-bank communications, economic releases, and news.

Generative AI may also expand access to research. Institutional-quality information can be translated into concise briefings, multilingual summaries, or explanations adapted to different levels of financial knowledge. This can help advisors, portfolio managers, and individual investors process a broader information set.

\subsection{Wealth Management and Personalized Financial Guidance}

Wealth-management applications center on personalization and communication \cite{challoumis2024ai}. A generative assistant can prepare meeting briefs, summarize prior client interactions, draft follow-up messages, explain portfolio performance, and generate educational material aligned with a client's goals. It may also help advisors explore scenarios such as retirement timing, cash-flow needs, tax-aware withdrawals, or changes in risk tolerance.

A tool-enabled system can combine client data with deterministic financial calculators. The language model interprets the question and communicates the result, while portfolio analytics, optimization software, or planning engines perform the numerical work. This division allows the system to provide natural interaction without treating free-form text generation as a substitute for financial calculation.

For self-directed users, conversational interfaces may make complex financial concepts more accessible. Potential uses include budgeting support, savings plans, explanation of investment products, and guided comparison of alternatives. The key innovation is not only automated advice, but the ability to adapt explanations and examples to the user's context.

\subsection{Customer Engagement and Service}

Financial customer service involves large volumes of repetitive but context-dependent questions \cite{kim2024assessing}. Generative AI can answer product questions, explain transactions, summarize account activity, assist with form completion, and draft responses to complaints or service requests. It can also provide real-time assistance to contact-center employees by retrieving relevant product terms and suggesting response language.

Compared with rule-based chatbots, generative systems can handle a wider range of phrasing and maintain context across follow-up questions. They can also summarize a conversation before handoff to a human representative, reducing the need for customers to repeat information. Multilingual generation may broaden service availability and reduce translation delays.

Beyond reactive support, models may generate personalized education, alerts, or next-step explanations. A banking application could explain why a payment is pending, summarize changes in monthly spending, or convert a complex fee schedule into a user-specific example.

\subsection{Lending and Credit Workflows}

Generative AI can support several stages of lending without serving as the underlying credit-scoring model. During application processing, multimodal systems can extract information from tax forms, bank statements, pay stubs, and business documents. They can summarize borrower narratives, identify missing information, and prepare structured case files for review.

For commercial lending, models can assist with financial spreading, covenant extraction, industry research, and credit memorandum drafting. They can compare borrower performance with peers, summarize management discussions, and organize evidence about repayment capacity. In consumer lending, potential uses include applicant assistance, document intake, plain-language explanations, and servicing communications \cite{jiang2025intellichain,golec2025interpretable,sanz2024credit}.

Generative systems can also support portfolio management after origination. They may summarize news about borrowers, interpret changes in financial statements, or draft monitoring notes based on newly available information. These applications combine document intelligence with ongoing knowledge synthesis.

\subsection{Risk Analysis, Fraud Investigation, and Financial Crime}

Risk functions depend on converting signals into explanations and actions. Generative AI can summarize market, credit, liquidity, and operational risk reports; explain changes in key metrics; and help users investigate unusual exposures. A tool-using assistant may query risk data, compare current values with historical ranges, and generate a narrative describing the main drivers \cite{tiwari2026generative}.

Fraud and anti-money-laundering teams face a related challenge: analysts must review transactions, customer profiles, alerts, and external information. Generative AI can organize this evidence into investigation summaries, create timelines, identify relationships among entities, and draft case narratives. It may also translate natural-language investigative questions into graph searches or database queries.

Cybersecurity applications include summarizing threat intelligence, explaining alerts, generating incident reports, and assisting with code or configuration analysis. Across these use cases, the model's advantage is its ability to combine heterogeneous evidence and produce an intelligible narrative for expert review.

\subsection{Operations, Reporting, and Document Processing}

Financial operations contain many document-heavy processes that are candidates for generative automation. Examples include payment exception handling, trade confirmation review, reconciliation commentary, invoice processing, onboarding, client documentation, and corporate-action processing. A Gen AI object can extract fields, interpret free-text instructions, classify documents, and prepare a proposed resolution \cite{pingili2025ai}.

Reporting is another broad area of use. Generative AI can create narrative commentary for financial statements, management dashboards, portfolio reports, and operational metrics. It can compare current and prior periods, highlight material changes, and produce audience-specific summaries. The same system may generate a brief executive overview and a more detailed analyst explanation from the same underlying data.

In legal and contract-related operations, models can extract clauses, compare agreements, identify nonstandard language, and prepare summaries for specialists. These functions are especially relevant to derivatives documentation, vendor contracts, loan agreements, insurance policies, and procurement.

\subsection{Software, Data, and Quantitative Productivity}

Software development is one of the earliest areas of generative AI adoption. In finance, coding assistants can generate data pipelines, SQL queries, unit tests, documentation, and examples. They can translate legacy code, explain unfamiliar systems, and help developers troubleshoot errors. Quantitative analysts may use models to prototype statistical tests, construct backtesting code, or convert mathematical ideas into executable scripts.

Generative AI can also make data platforms easier to use. A natural-language interface can translate a business question into a database query, generate a chart, and explain the result. This lowers the barrier between financial domain experts and technical systems. The resulting productivity gains may be substantial because many financial workflows require repeated movement between business language, data definitions, code, and presentation.

Synthetic data generation is another possible application \cite{lee2024comprehensive}. Models can produce realistic but artificial text, transactions, or documents for software testing, scenario exploration, and model development. In quantitative finance, generative methods may also be used to simulate market paths or rare-event scenarios, complementing traditional stochastic approaches.

\subsection{Insurance and Claims Services}

Insurance combines complex products, long documents, customer interaction, and event-driven workflows. Generative AI can summarize policies, explain coverage, assist with underwriting submissions, extract information from claims documents, and draft claim correspondence. Multimodal models may interpret photographs, repair estimates, medical forms, or scanned records together with textual descriptions \cite{amistapuram2025generative}.

For underwriting, a model can compile applicant information and external research into a structured summary. For claims, it can create a chronology, identify missing documents, compare the claim with policy language, and prepare a case brief. Customer-facing applications can explain deductibles, exclusions, and next steps in more accessible language.

\subsection{Cross-Functional Application Portfolio}

Table~\ref{tab:use_cases} consolidates representative applications and the primary capability each one uses.

\begin{longtable}{p{0.20\textwidth} p{0.29\textwidth} p{0.21\textwidth} p{0.20\textwidth}}
\caption{Representative generative AI use cases in finance}\label{tab:use_cases}\\
\toprule
\textbf{Financial function} & \textbf{Illustrative use case} & \textbf{Primary capability} & \textbf{Typical output} \\
\midrule
\endfirsthead
\toprule
\textbf{Financial function} & \textbf{Illustrative use case} & \textbf{Primary capability} & \textbf{Typical output} \\
\midrule
\endhead
Investment research & Earnings and filing synthesis & Knowledge synthesis & Research brief and key themes \\
\addlinespace
Capital markets & Market-event and news analysis & Synthesis and analytical assistance & Event summary and scenario narrative \\
\addlinespace
Wealth management & Advisor meeting preparation & Personalization and generation & Client brief and talking points \\
\addlinespace
Customer service & Conversational product support & Interactive assistance & Contextual answer or handoff summary \\
\addlinespace
Lending & Credit memorandum drafting & Document intelligence and generation & Structured borrower analysis \\
\addlinespace
Fraud and AML & Alert investigation support & Workflow orchestration & Evidence timeline and case narrative \\
\addlinespace
Operations & Exception handling & Multimodal extraction and tools & Proposed resolution and documentation \\
\addlinespace
Finance and reporting & Variance commentary & Analytical assistance and generation & Management narrative \\
\addlinespace
Technology & Code and test generation & Coding assistance & SQL, Python, tests, and documentation \\
\addlinespace
Insurance & Claims-file summarization & Multimodal synthesis & Claim chronology and case brief \\
\bottomrule
\end{longtable}

\subsection{Sources of Business Value}

The business value of generative AI is likely to arise through several channels. First, it can reduce search and reading time by compressing large information sets into task-specific summaries. Second, it can accelerate document and communication production by creating high-quality first drafts. Third, it can broaden access to analytical tools by allowing users to interact through natural language. Fourth, it can improve continuity across workflows by carrying context from research and calculation into documentation and communication. Finally, it can enable new products built around personalized, conversational, and multimodal interaction.

Value will vary by task. High-volume and document-intensive activities may produce immediate efficiency gains, while research and advisory applications may create value by expanding the information considered or improving the speed of iteration. Agentic systems may eventually automate longer process segments, but near-term adoption is likely to emphasize copilots that augment domain experts.

\subsection{Technical and Research Challenges}

Although this paper emphasizes applications rather than governance, several technical limitations remain central to future development. Generative models can produce unsupported statements or fabricated citations \cite{ji2023hallucination,maynez2020faithfulness}. Performance may degrade when documents are long, retrieval is incomplete, or financial terminology is highly specialized. Quantitative reasoning can also be unreliable when a model is asked to calculate without an external tool.

Research is therefore needed on finance-specific benchmarks, grounded generation, temporal reasoning, numerical accuracy, multimodal document understanding, and evaluation of agentic workflows. Another open question is how best to combine language models with established financial models. Hybrid systems that connect generative interfaces to deterministic analytics, forecasting models, and domain databases are likely to be more useful than models operating in isolation.

%% file: NewSections/04_Conclusion.tex
\section{Conclusion}
\label{sec:conclusion}

Generative AI introduces a new interaction layer across financial data, documents, software, and human expertise. Its potential in finance is broader than conversational customer service. The same core technologies can support investment research, wealth management, lending, fraud investigation, operations, reporting, insurance, and software development. Their common contribution is the ability to transform unstructured information into searchable knowledge, analytical support, personalized communication, and multi-step digital workflows.

The most promising applications combine generative models with domain-specific data and specialized tools. Retrieval systems provide access to current financial documents, analytical engines perform reliable calculations, multimodal models interpret complex forms and reports, and agentic components coordinate tasks across systems. This hybrid architecture allows generative AI to complement rather than replace established financial analytics.

Future research should focus on finance-specific evaluation datasets, numerical and temporal reasoning, multimodal document intelligence, personalized interaction, and the design of effective human--AI collaboration. As these capabilities improve, generative AI may become a general-purpose productivity and interaction layer for financial institutions, reshaping how financial information is analyzed, communicated, and converted into action.